# Application of laser-induced breakdown spectroscopy and neural networks on archaeological human bones for the discrimination of distinct individuals

Panagiotis Siozos[1], Niklas Hausmann[2,3], Malin Holst[3,4], Demetrios Anglos[1,5]

[1] *Institute of Electronic Structure and Laser, Foundation for Research and Technology-Hellas (IESL-FORTH), P.O. Box 1385, GR 711 10, Heraklion, Crete, Greece*

[2] *Niedersächsisches Institut für Historische Küstenforschung, Viktoriastr. 26-28, 26382 Wilhelmshaven, Germany*

[3] *Department of Archaeology, University of York, Kings Manor, Exhibition Square, YO1 7EP York, UK*

[4] *York Osteoarchaeology Ltd, 75 Main Street, Bishop Wilton, York, YO42 1SR*

[5] *Department of Chemistry, University of Crete, P.O. Box 2208, GR 710 03, Heraklion, Crete, Greece*

*Corresponding authors' e-mail addresses: psiozos@iesl.forth.gr, niklas@palaeo.eu*

**Abstract**

The use of elemental analysis based on Laser-Induced Breakdown Spectroscopy (LIBS) combined with Neural Networks (NN) is being evaluated as a method for assigning archaeological bone remains to individuals. The bone samples examined originate from excavations of burials at the Cross Street Unitarian Chapel, Manchester (United Kingdom) tha date from the 17[th] to the 19[th] century. In this study, we critically assess the influence of soil contaminants, by separating the bone elemental fingerprint into two groups of different components prior to the NN analysis. The first group includes elements related to the bone matrix (Ca and P) as well as elements that are regularly incorporated in the living bone tissues (Mg, Na, Sr, and Ba). The second group includes metals with a low probability of accumulation in living bone tissues whose presence is more likely to be related to diagenesis and the chemical composition of the burial soil (Al, Fe, Mn). The NN analysis of the spectral data, based on the use of an open access software, provided accurate results, indicating that it can be a promising tool for enhancing LIBS applications in osteoarchaeology. The influence of bone diagenesis and soil contaminants is significant. False classifications occurred exclusively in the NN analyses that relied partially on elemental peaks from the second group of elements. Overall, the present study indicates that discrimination between individuals through LIBS and NN analysis of bone material in an archaeological setting is possible, but a targeted approach based on selected elements is required and the influence of bone diagenesis will have to be assessed on a case-by-case basis. The proposed LIBS-NN method has potential as a tool capable for distinguishing distinct individuals in disarticulated or commingled human skeletal assemblages particularly if combined with standard osteometric methods.

**Introduction**

The excavation and methodical examination of graves with disarticulated, commingled, multiple individuals or mass graves poses a major problem to archaeologists and anthropologists given that a key and demanding aspect of their analysis is the separation and grouping of commingled human remains into individual skeletons (Caffell, 2009; Caffell, 2012; Nikita, 2017). A reliable method to make these associations provides researchers the means to make accurate osteological or palaeodemographic interpretations (Adams, 2014) but becomes all the more difficult as the size of the skeletal assemblage increases or when commingling is extensive (Adams, 2008; Kendell, 2014).

Currently, several well-established methods (macroscopic and non-destructive) are being extensively used in re-associating commingled skeletal remains. These are primarily based on osteometrics but often utilize other aspects of skeletal variation, for example, pathology, age, sex and activity markers (Byrd and Adams 2003; Byrd, 2008; Byrd and LeGarde 2014; Nikita and Lahr, 2011; Thomas, 2013).

Modern techniques for the analysis archaeological DNA (aDNA) are increasingly deployed in commingled assemblages, to identify fragmentary evidence and assemble separated remains of single individuals following mass disasters (Adams, 2014). However, even this powerful tool suffers several drawbacks related to contamination with modern genetic material coming from sample handling or the poor preservation of aDNA in many skeletal parts (with the exception of ear ossicles or petrous bone). Moreover, DNA analysis, when studies of commingled assemblages are considered, would require sampling and destructive analysis of a very large number of skeletal remains, increasing costs enormously but still affording ambiguous results. Therefore, there is a strong interest for rapid and cost-effective methods that can be applied in large numbers of skeletal remains and ideally on-site.

Recently, elemental fingerprinting through Laser Induced Breakdown Spectroscopy (LIBS) combined with a supervised Neural Networks (NN) algorithm was applied by Moncayo *et al.* (Moncayo, 2014), who achieved fast and robust discrimination of human remains from modern individuals, based on the characteristic elemental fingerprints of their bones and teeth. The proposed method could be considered as a simple and cost-effective alternative for the slower and more expensive biomolecular methods to discriminate between modern human remains or even animal remains (Marin-Roldán, 2020) and could be applied as a screening tool in the context of forensic analysis of victims from natural disasters, in crime scenes or in relation to genocide acts.

There is additional potential to apply this type of screening method to human remains in archaeological contexts, similar to studies applying elemental fingerprinting and neural networks to archaeological material culture, for instance ancient or historical pottery (Botto, 2019; D'Andrea, 2014; El Haddad, 2013; Qi, 2018; He, 2019; Ramil, 2008). However, applying elemental fingerprinting in an osteoarchaeological context requires certain additional considerations. In contrast to ceramic materials, human bones undergo significant taphonomic changes in their burial environment (Kasem, 2011; Hedges, 2002). Bone diagenesis may involve dissolution of the bone tissue or cementation of it by extraneous minerals, and recrystallization of the bone mineral or its replacement by other mineral species. All these diagenetic processes cause changes in the elemental composition of the bones. Therefore, the basic requirement for the interpretation of the elemental composition in a buried bone is the identification of the original biogenic components and their concentration with negligible or limited interference from the diagenetic elements.

The main objective of the present research work is thus to investigate the use of the NN method applied on LIBS spectra of archaeological bone remains as regards its potential to aid discrimination of individuals and further to examine limitations related to the diagenetic influence of the burial environment (i.e. soil) on the elemental composition of the bones.

**Materials and methods**

*Osteological material*

Seven different bones from five individuals were used in this study (Fig. 1 and Table 1). These samples originate from archaeological excavations of burials at the Cross Street Unitarian Chapel, Manchester (United Kingdom). The chapel was in use from 1694 to 1852 but only few burials could be associated with a more accurate date, making it difficult to reliably date our sample material. The burial register covers a narrower time range 1785–1840, but gravestone inscriptions for several individuals (including individual SK 3.50 of this study) predate the burial register, indicating that the period of the chapel's use would be a more truthful estimate of the sample ages, where gravestone inscriptions are not available (Keefe, 2017). Although not always preserved, wooden coffins were used when burying humans. Nevertheless, the remains varied in completeness and preservation. The skeletons selected for the present analysis were at least 50% complete and their state of preservation ranged from 'good' to 'extremely poor' in the case of SK 3.50 (*ibid*). To gain sample material we retrieved fragments of approximately 5 cm in size using bones which had already been cleaned with water and had been air-dried immediately following their excavation in 2015. Bone fragments were stored in plastic bags until the spectroscopic analysis was carried out.

*LIBS analysis and setup*

Laser Induced Breakdown Spectroscopy (LIBS) is a spectrochemical method that provides information about the elemental composition of materials. Focusing a pulse from a nanosecond laser onto the surface of an object or sample results in material ablation and formation of a transient micro-plasma, which emits light upon its relaxation. The plasma emission is spectrally analysed and recorded on a spectrometer thus producing a LIBS spectrum, which features intense, sharp atomic emission peaks, used to identify the elements contained in the sample on the basis of their characteristic wavelength. The emission intensity at the maximum of a spectral peak or, alternatively, the total (or integrated) emission corresponding to a peak is related to the number density of each emitting species in the plasma plume and this, in turn, with the concentration of specific elements in the ablated material (Anglos, 2001; Anglos and Detalle, 2014; Anglos, 2019; Botto, 2019; Lee, 2004). Considering that laser ablation is involved in a LIBS measurement, removal of a minute amount of material (on the order of a few nanograms) from the sample surface takes place with each laser pulse.

The speed of LIBS analysis permits handling of large numbers of samples in short times, while its portability makes it a user-friendly and flexible tool for materials analysis in many different, sometimes difficult-to-access, locations. Several LIBS spectrometers have been specifically developed to be compact and portable (Anglos 2014) with the aim to support field expeditions including archaeological campaigns or studies at historical or archaeological museums (Westlake, 2012; Papliaka, 2016; Baker, 2017; Siozos, 2017; Pérez-Diez, 2020).

The LIBS system used in this study has been presented elsewhere (Hausmann, 2017; Hausmann, 2019a; Hausmann, 2019b). It makes use of an infrared (λ = 1064 nm) Q-switched Nd:YAG laser (SL-404, Spectron Laser Systems), that emits pulses of 10 ns duration, which were focused directly onto the sample surface using an objective lens (10x magnification, focal length, f = +28 mm, LMH-10X-1064, Thorlabs) with infrared anti-reflection coating. Each pulse has energy $E_P \approx 10$ mJ and irradiates the sample on the exposed surface in a circular area of approximately 50 μm in diameter. Within this area, the laser irradiation creates a luminous plasma plume of partly ionised sample material. A quartz fibre collects the light emitted by the plume and guides it into a compact Czerny-Turner spectrometer (Avaspec-2048-USB2, Avantes) that was used to record the spectra in the range of 200 - 466 nm with a resolution of 0.2 nm. The delay time applied on the CCD was $\tau_D = 1.28$ μs and the integration or gate time was $\tau_G = 1$ ms. It is noted here that spectra are typically recorded with a short time delay, $\tau_D$, with respect to the initiation of the plasma formation, as this results in cleaner and better-resolved spectral features. Fifty-five to sixty different points were analysed on each bone by LIBS. Ten laser pulses were applied, at a repetition rate at 0.5 Hz, on each point. The first 5 pulses were used to remove superficial dirt and dust revealing a clean surface, while the next 5 pulses in the set produced an averaged LIBS spectrum per each sampling point. These were the data subjected to Principal Component Analysis (PCA) and the NN processing for discriminating the bones.

The discrimination of the biogenic from the diagenetic signal is rather difficult and requires complex methods applied mainly in palaeodietary research (Makarewicz and Sealy, 2015). In the present case, we have applied and evaluated a simple approach that potentially can be used for the discrimination of individuals and is based on distinguishing two groups of elemental peaks in the LIBS spectra (Table 2), which are then used independently as an input to the NN model for training and testing.

The first group (group 1) includes calcium (Ca) and phosphorus (P), since these elements are related to the intrinsic chemical composition of the bones, namely hydroxyapatite, ($Ca_{10}(PO_4)_6(OH)_2$) and calcium phosphate, $Ca_3(PO_4)_2$. Additionally, group 1 includes elements such as magnesium (Mg), sodium (Na), strontium (Sr) and barium (Ba), soluble in the blood, that can be easily deposited in living bones and therefore can be used with higher reliability for discriminating among individuals (Priest, 1990). The second group (group 2) corresponds to metals with low solubility in the blood, in particular iron (Fe), aluminium (Al) and manganese (Mn). These metals have been repeatedly correlated with diagenetic processes (Kuczumow, 2010; Lambert, 1982; López-González, 2006; Rasmussen, 2013; Rasmussen, 2019) given their abundance in soil and groundwater. Two additional elements, copper (Cu) and lead (Pb), have been detected in the LIBS spectra but were excluded from group 2 despite their low solubility. Concerning copper, there exists a debate about the influence of diagenesis on it with some studies reporting it to be unaffected by diagenesis (Byrne, 1987; Lambert, 1979) and others arguing for the opposite (Lambert, 1984). The content of lead has been reported to be affected by postmortem alterations (Wittmers, 2008), however, given that lead can also be detected if an individual consumed food or inhaled air with a high lead concentration for a long period of time it can equally well be incorporated into the bone structure (Rasmussen, 2017).

To illustrate the positions (wavelength) of spectral peaks corresponding to elements in group 1 and group 2, we indicate in Fig. 2 selected regions in a typical LIBS spectrum recorded from the SK 3.25 femur sample. Not surprisingly, the most intense emission lines can be observed for singly ionized calcium atoms (Ca II) (Fig. 2d), reflecting both the high content of bone in hydroxyapatite and calcium phosphate and the high sensitivity of LIBS to Ca detection. Even though there exist several additional emission lines in the spectrum, mainly corresponding to Ca, we have restricted our focus on the spectral regions shown in Figure 2. These, on the one hand provide sufficient information about the elemental composition of the samples, representative of both the bone matrix and the burial background, and on the other hand keep the dataset down to a minimum size reducing computer process time and avoiding overfitting.

In contrast to the emission of calcium, the signal from phosphorus, as recorded through transition lines at 213 – 215 nm, is considerably weaker (Fig. 2a) due to the very low sensitivity of our detector in the relevant spectral range (200–250 nm) and the high energy of the excited state of the transition (7 to 8 eV). Still, these lines adequately reflect the phosphorous content in the material (Li, 2014). Additional phosphorous lines are present in the 252-256 nm spectral range but overlap with emission lines of iron (Kondo, 2009). While the iron content of bones is relatively low, it can vary significantly given it is a diagenesis product accumulating in the bones from the soil in the burial area. It is further noted that a number of intense emission lines can be observed in the vacuum ultraviolet (VUV) spectral range, namely at wavelengths lower than 200 nm. Although, the VUV lines, observed between 165 and 190 nm, would be most suitable for phosphorous detection, the strong absorption from atmospheric air in this spectral range (Marín-Roldán, 2019) adds complexity to measurements given that one would need to maintain the spectrometer system and the light collection pathway either in vacuum or purged with an inert gas.

Finally, other intense peaks originating from elements in group 2, including Ba, Sr, Pb, Mn and Fe, are indicated in the spectra shown in Figure 2b and 2c.

*PCA data filtering*

Initially, Principal Components Analysis (PCA) was applied separately to the dataset corresponding to each bone (54–58 spectra) in order to detect any outlier spectra. Including outlier spectra into the NN model interferes with the discrimination ability of the model, therefore, spotting outliers by use of PCA and filtering these out is an effective way of improving the subsequent analysis in the NN model. Indeed, by using PCA as a means to remove outliers, we were able to filter out inconsistencies within the datasets for each individual bone sample. As a result of this filtering procedure the number of spectra for each bone was reduced from 54–58 down to 50. Figure 3 illustrates as an example the result of PCA filtering in the case of sample SK3.25-sca.

*Neural Network algorithm*

The open-access R-package "neuralnet" (Version 1.44.2) was used to perform the NN analysis (Fritsch, 2019). R is a programming language appropriate for statistical computing and graphics which provides a wide range of statistical and graphical tools (R Core Team, 2019). The advantage of R is its availability as a free, open software under the terms of the Free Software Foundation GNU General Public License in source code form. Currently, a vast section of academia uses R language and contributes to the development of packages, allowing, for example, the implementation of specialised algorithms or techniques, and expanding enormously its capabilities beyond any commercial statistical software. Due to this, the method proposed herein can be easily adapted from experts in the field of osteoarchaeology and forensics, creating a positive impact on their work (Tippmann, 2015). The 'neuralnet' package was obtained from the CRAN repository and applied to develop and construct the corresponding NN models in R (Version 3.6.2). The globally convergent algorithm of 'neuralnet' is based on the resilient back-propagation without weight back-tracking (Fritsch, 2019). This particular package was selected on the basis of its frequent use by experts in biology and other scientific disciplines as well as its continuous maintenance and testing.

The neurons in the input layer comprised the emission intensity values at the selected spectral regions as shown in Figure 2. In the present work, 157 neurons were considered as the input layer corresponding to the group 1 dataset, 185 neurons for the group 2 dataset and the sum of the two, 342 neurons, collectively for groups 1 and 2. The number of input-neurons corresponds to the number of pixels (data points) across the selected spectral windows. A single layer of 10 neurons was considered in the hidden layer and the number of individuals (5) in the output layers.

Neural network classifications can be evaluated by removing some of the data (in our case spectra) and training the network on the remaining subset. Following the training stage, those spectra that were previously withheld were used as a test for the model. In the present case, 300 spectra out of a total of 350 (86%) were selected randomly to train the network and the remaining 50 spectra (14%) were used to test the network.

Four different preprocessing methods were applied for standardizing the spectral data in both the training and the test sets and their influence on the robustness of the model was examined. These were: 1) normalization by the sum of the intensity values of each spectrum, 2) normalization to the square root of the sum of the squared intensity values of the spectrum, 3) normalization by the maximum intensity value of each spectrum and 4) normalization to the variance of the dataset. The percent of success in the tested dataset of group 1 for all individuals was calculated and the process time required for the NN method to converge was measured for each normalization method as shown in Table 3. All normalization methods had a success rate higher than 92%. The two most effective ones were the sum of intensities method (2) and the unit vector normalization method (3) with a success rate of 96.4% and 96.2% respectively. However, there is a significant difference in the processing time with the unit vector requiring almost half of the time to converge in comparison to the sum of intensities method. We thus selected the unit vector method to be used for the NN analysis.

The accuracy of the neural network analysis was estimated by applying the "spectral correlation" (SC) parameter proposed by Moncayo *et al* in several of their papers (Moncayo, 2014; Moncayo, 2015; Moncayo, 2017). The parameter, defined as the percentage of test spectra correctly classified, was assessed as model accuracy on the results of the test dataset.

$$SC = \frac{100}{N} \sum_{i=1}^{N} d_i \qquad (1)$$

where $d_i$ is either 1 when a spectrum is classified correctly or 0 otherwise, and N is the total number of spectra.

To consider a bone sample as being accurately classified, the predicted class of the model must agree with the actual class when the spectral correlation parameter exceeds a threshold (arbitrarily set at) $SC_{th} \geq 80\%$. A bone sample is defined as 'unassigned' when $SC < SC_{th}$ for either one of the five individuals.

Given these preconditions, we performed several neural network analyses, one that included all elements, and two that distinguished between elements falling into the 'biogenic' and 'diagenetic' groups to compare their varying results. Finally, we simulated a real case scenario to evaluate the performance of the neural network model and the overall capability of it for associating separated bones from the same individual under realistic analysis conditions. In this experiment, the training and test spectra were not selected randomly and across all bones, but instead 250 (71%) spectra from five bones - each belonging to one of the five individuals - were used to train the network and 100 (29%) spectra of two bones that belong to individuals 1 and 2 (SK 3.25 scapula and SK 3.50 clavicle) were used for testing.

**Results and Discussion**

Initially, PCA was applied to the data obtained from all bone samples, first separately on the selected spectral windows related to the bone matrix components (group 1) and to the soil background (group 2) and then to both of them together. The aim of this procedure was to assess the capacity of PCA to provide any kind of discrimination of individuals (Fig. 4). While spectra appear to largely cluster by individual, any discrimination between individuals based on the PCA score plots is rather unreliable due to the large overlap among the clusters. It is also worth noting that the clusters for individuals 1 and 2 are the least concentrated, most likely due to the fact that these individuals were represented by two rather than one bone sample. In fact, individual 1 is represented with two clusters, which are quite distinct in the PCA analysis of spectra reflecting group 2 elements (Fig. 4b).

The process of training and testing the neural network was performed in less than a minute when the code was executed on a desktop computer using the datasets of groups 1 and 2 (157 and 185 input neurons, respectively). The process time increased by a factor of five when the same code was executed by including the dataset of the whole spectral range (2048 pixels or input neurons). It is also clear that processing time will increase exponentially if the number of datasets from skeletal remains is increased.

The results of the initial neural network analysis including both groups to train and test the NN model are promising as shown in Table 4, with SC values classifying all the datasets to the correct individual with a high percentage. The neural network model based solely on the group 1 dataset also discriminates correctly all seven bones (Table 4) with respect to the five individuals based on the SC values, with 100% certainty for all 5 individuals. On the other hand, with the NN model relying only on the group 2 dataset, six bones were assigned correctly to the five individuals while one bone was not assigned to any one of them (Table 4) since the corresponding SC value was 73%, i.e. below the confidence threshold. Generally, the SC values that resulted based on the group 2 datasets were all lower compared to the corresponding ones for group 1.

*Real case scenario*

As already mentioned, in the simulation of a real case scenario, a non-randomized selection of training and test spectra is made, with two bone samples excluded entirely from the neural network training, as would be the case in an actual situation, in which one would try to associate bones of unknown relation to a range of other bones attributed with certainty to different individuals. The results of the discrimination based on group 1, group 2, and the combined group 1 and 2 datasets are presented in Table 5.

Using the group 1 dataset, the NN model correctly assigns the SK 3.50 clavicle bone to individual number 2 with an SC value of 91%. However, it fails to classify the SK 3.25 scapula bone to the correct individual (4 instead of 1). Since the SC (53%) for individual 4 is below the threshold value of 80%, this bone is classed as 'unassigned' due to low confidence in this result.

Using the group 2 dataset the NN model failed to classify the SK 3.25 scapula bone to its correct individual (4 instead of 1), however, since the SC (72%) was again below the confidence value of 80%, this bone remained as 'unassigned'. Moreover, the NN model only achieved an SC value of 69% for the SK 350 clavicle bone, leaving the sample, too, 'unassigned'.

The results were mixed when using the combined elemental datasets of group 1 and 2. Once again the SK 3.50 clavicle bone was assigned correctly to individual number 2 based on a SC value of 82%. However, the model provided a false positive classification for the SK 3.25 scapula bone, which was assigned to individual 3 with a high confidence (SC = 90%).

The results of this study indicate that the elemental peaks related to the bone matrix composition can lead more reliably to bone discrimination while those originating from the soil background were inadequate for providing accurate bone discrimination using NN models on bone datasets that were selected for testing. However, the probability to acquire false positives turned out to be higher in the case of elemental peaks related to both bone and soil chemical composition, compared to elemental peaks related only to the bone matrix. The failure to assign the dataset from bone sample SK 3.25 scapula to individual 2 can be attributed to extensive contamination and diagenesis of the bone that has given rise to additional spectral lines interfering with the elemental peaks of species related to the bone matrix. It is conceivable that these errors could be mitigated by an increased number of bones per individual to train the NN, rather than relying only on a single bone for training. It is also possible to take the specific composition of the local soil into account to understand the origin of some elements in the bone and use this understanding to quantify their influence on the NN model. Detailed information on the burial soil composition is not available for the skeletal remains in our study. Nevertheless, this information could be helpful when considering the potential for a difference in diagenetic influences within the assemblage and to prevent false positives driven by the seemingly more influential soil-related elements. As such, the effect of soil composition on the elemental spectrum should be considered and quantified in future studies.

Lastly, it was noticed that emission peaks from copper (Cu) differ strongly between individuals, with males indicating higher Cu concentrations than females within our test group. A clear discrimination by sex was observed by comparing the intensity ratio of the Cu emission peak at 324.754 nm against the Ca peak at 315.887 nm (Fig. 5). This pattern could indicate that individual 4, whose sex was unidentified, was female due to the low Cu/Ca ratio. However, there is not sufficient information published, to our knowledge, regarding any difference in the accumulation of Cu in the bones between males and females. That being said, other studies did find that the immediate environment during the individual's lifetime can contribute to such concentration contrasts for certain elements (Rasmussen et al. 2013; 2017; 2019; 2020), suggesting that there might be sex-dependent exposure to the element as a result of gendered living spaces or working conditions.

**Conclusions**

A method based on Laser-Induced Breakdown Spectroscopy (LIBS) combined with a supervised Neural Networks (NN) algorithm has been applied for the first time on archaeological bones to discriminate human remains based on characteristic spectral features and to reveal the influence of chemical exchange with soils in archaeological material.

The LIBS spectra were divided into two groups. The first group includes the two main elements related to the chemical composition of the bone matrix (Ca and P) and in addition those that can be easily deposited in living bone tissues (Na, Mg, Sr and Ba) and therefore can be used with higher reliability for the discrimination of individuals. The second group corresponds to metals with a smaller accumulation probability on living bone tissues (Al, Fe, Mn) and therefore their presence is related to diagenesis thus contamination of the bones from the soil of the burial area. The biogenesis group provided partly successful discrimination in a realistic scenario, while the diagenesis group failed to provide any kind of successful discrimination. It was further found that elemental peaks related to bone matrix and burial soil composition, considered jointly, have a high probability of giving rise to false positives compared to elemental peaks related only to the bone matrix.

The open-source NN software (R package 'neuralnet') demonstrated its potential to be a reliable tool for LIBS applications. It has several advantages as it continues to evolve, has reduced costs, and exhibits a high adaptability and interoperability with respect to existing requirements. Therefore, the use of open-source NN software and tools can potentially boost the application of advanced machine learning methods in the particular field of LIBS analysis.

Overall, our results indicate that the elemental peaks related to the bone matrix composition can lead to fast discrimination of individuals with the potential to be applied cost-effectively and while in the field. Important parameters need to be further explored which are relevant to the optimum interaction of the laser beam with the bone matrix coupled to efficient collection of the emission signals, selection of clean spectral lines free from interferences, particularly those arising from the diagenetic background, and proper sampling across the bone surface. Moreover, more detailed empirical data on how to mitigate the influence of diagenesis and avoid false-positives is required before confident discrimination between individuals can be achieved. An additional factor to be taken into account relates to the fact that LIBS is a minimally (or micro-) destructive method thus obtaining official permit for sampling of each sample will always be necessary.

Finally, combination of the LIBS-NN approach with osteometric or other macroscopic examination and sorting methods, extensively applied in osteoarcheological studies, needs to be considered. This is expected to generate even more reliable results exploiting skeletal parts, already assigned to an individual on osteometric grounds, for training and testing the NN models or just bypassing their analysis and thus reducing analysis time.

**Acknowledgments**

We would like to express our gratitude to Dr S. Moncayo for his useful advice related to the application of neural network analysis on LIBS spectra and to Dr A. Nafplioti for helpful discussions. Moreover we are grateful to CFA Archaeology for providing the bone samples examined in this study.

This research was funded in part by the European Commission H-2020 Program through the Marie Skłodowska-Curie Individual Fellowship project 'ACCELERATE' (Grant Agreement No. 703625). We also acknowledge support by projects "HELLAS-CH" (MIS 5002735) and POLITEIA II (MIS 5002478), both implemented under the "Action for the Strategic Development on the Research and Technological Sector," funded by the Operational Programme "Competitiveness,

Entrepreneurship and Innovation" (NSRF 2014-2020) and co-financed by Greece and the European Union (European Regional Development Fund).

**Figure Captions**

Fig. 1: Images of the bone fragments: a) SK 3.25 Femur, b) SK 3.25 Scapula, c) SK 3.50 Clavicle, d) SK 3.50 Femur, e) SK 4.36 Ilium, f) SK 4.37 Tibia, g) SK 4.45 Humerus.

Fig. 2: Selected regions of a typical LIBS spectrum obtained from an archaeological bone in order of increasing wavelength (x-axis). The elemental peaks related to the structure of the bones (biogenesis) are indicated in blue (group 1) and those related to soil composition (diagenesis) are shown in magenta (group 2).

Fig. 3: PCA analysis of the SK 3.25 clavicle LIBS dataset for detection of outlier spectra via a multivariate t-distribution (dashed line).

Fig. 4: PCA analysis of the LIBS spectra using spectral peaks corresponding to elements in a) group 1, b) group 2 and c) both group 1 and group 2.

Fig. 5: Variation of Cu concentration in bones corresponding to identified sexes. a) Emission peaks of Cu normalised with respect to the Ca peak at 315.887 nm, in spectra recorded from the female bones (green), male bones (blue), and the bone of individual SK4.37 whose biological sex is unknown (black). b) Mean value of the Cu (324.754 nm)/Ca (315.887 nm) peak intensity ratio for each bone. Note that individual SK4.37 is within the range of the individuals identified as female.

**Table titles**

Table 1: List of individuals and bone samples examined in this study.

Table 2: Atomic emission lines in the LIBS spectra shown in Figure 2.

Table 3: Data normalization methods tested in the present work.

Table 4: Spectrum correlation (SC) values derived from NN analysis of LIBS spectral data.

Table 5: NN-LIBS analysis of spectral data for assigning two "unknown" bones. [a]

**Figures**

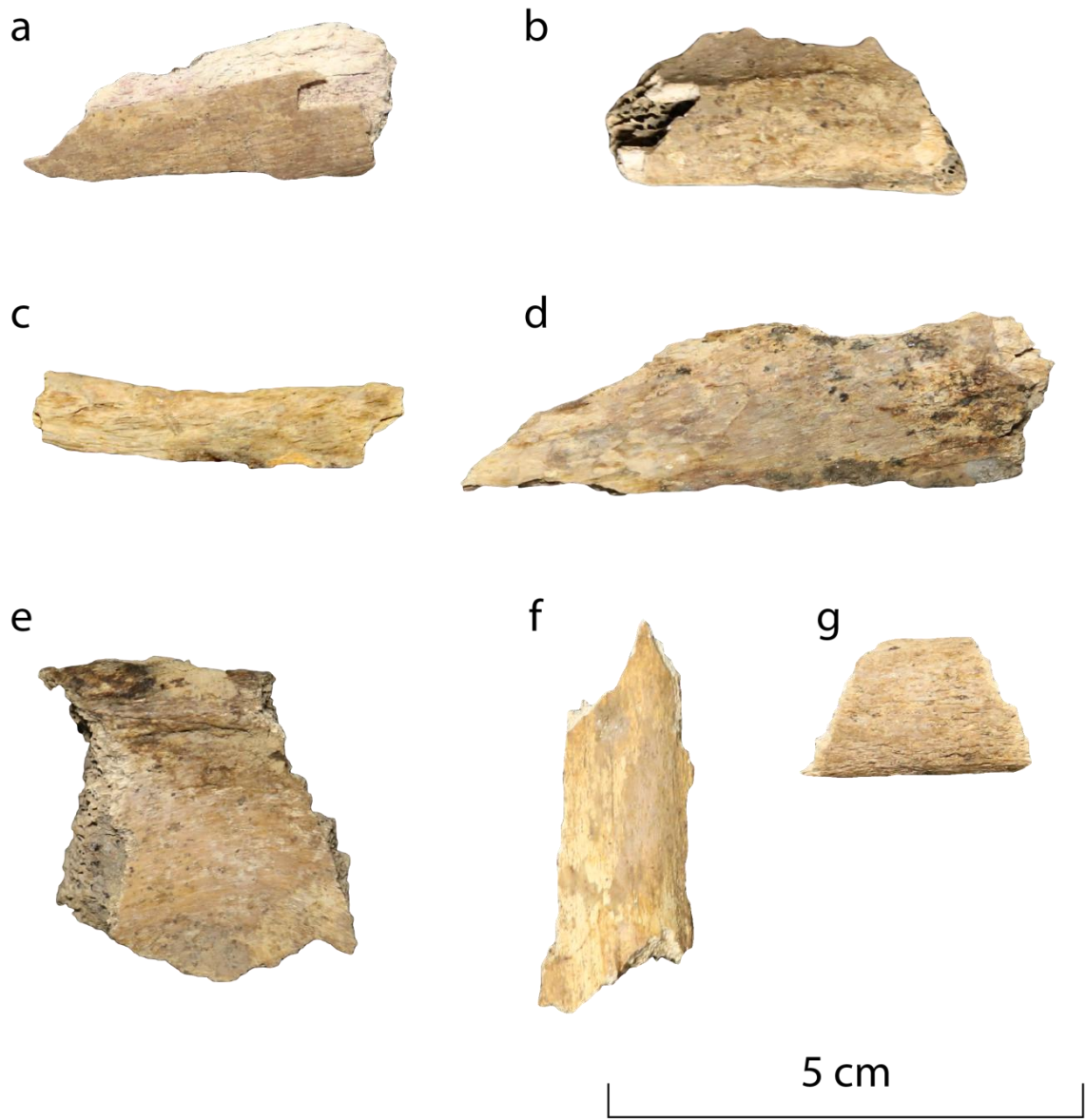

Fig. 1: Images of the bone fragments: a) SK 3.25 Femur, b) SK 3.25 Scapula, c) SK 3.50 Clavicle, d) SK 3.50 Femur, e) SK 4.36 Ilium, f) SK 4.37 Tibia, g) SK 4.45 Humerus.

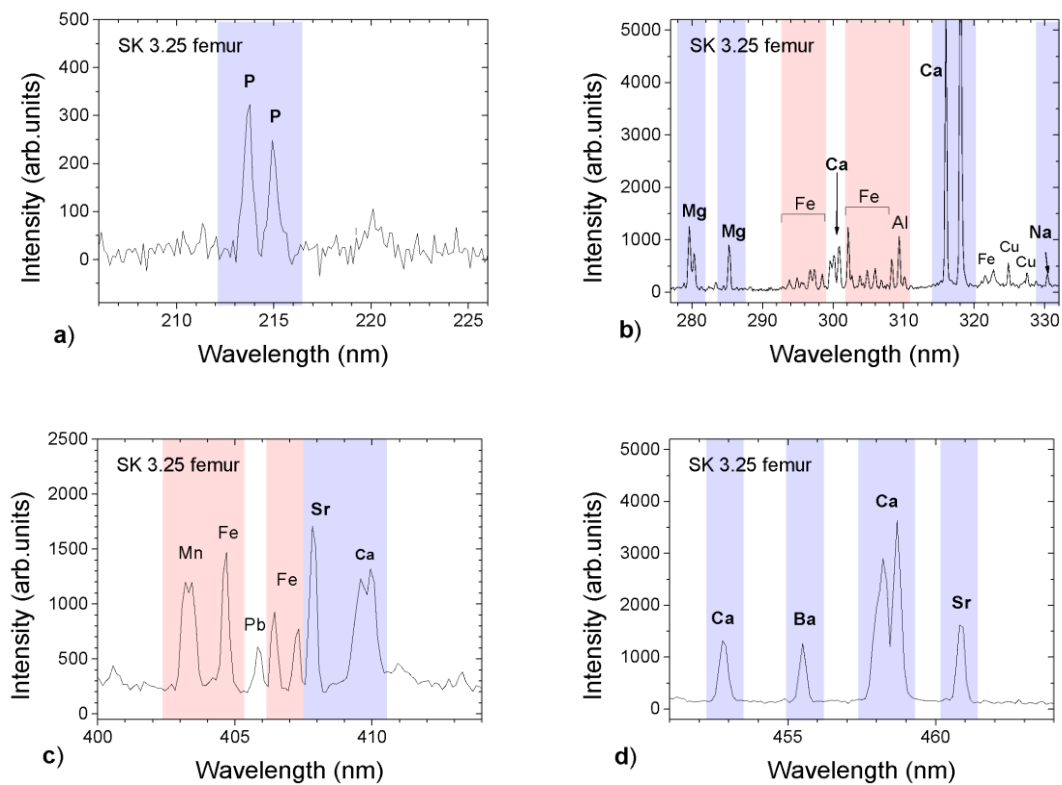

Fig. 2: Selected regions of a typical LIBS spectrum obtained from an archaeological bone in order of increasing wavelength (x-axis). The elemental peaks related to the structure of the bones (biogenesis) are indicated in blue (group 1) and those related to soil composition (diagenesis) are shown in magenta (group 2).

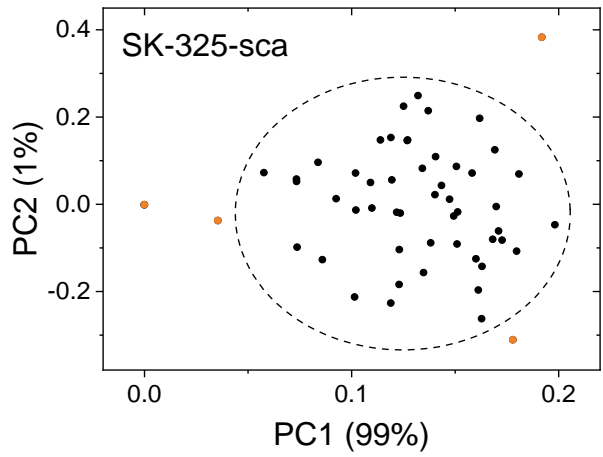

Fig. 3: PCA analysis of the SK 3.25 clavicle LIBS datasets for detection of outlier spectra via a multivariate t-distribution (dashed line).

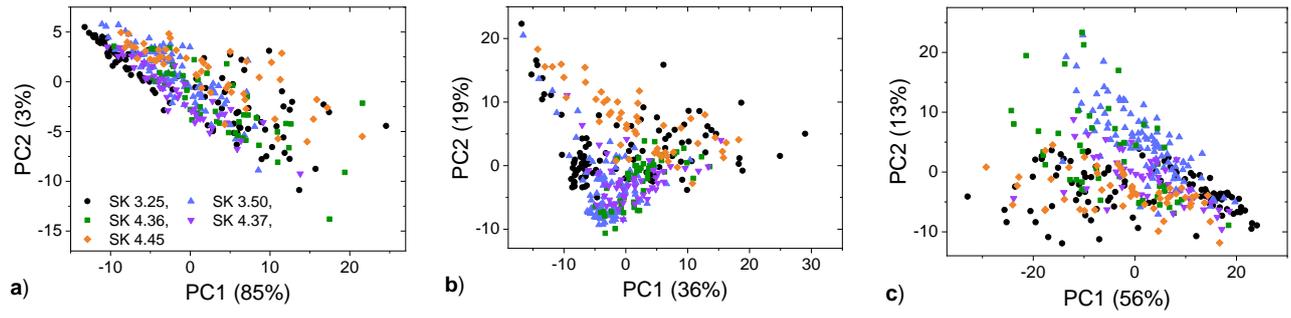

Fig. 4: PCA analysis of the LIBS spectra using spectral peaks corresponding to elements in a) group 1, b) group 2 and c) both group 1 and group 2.

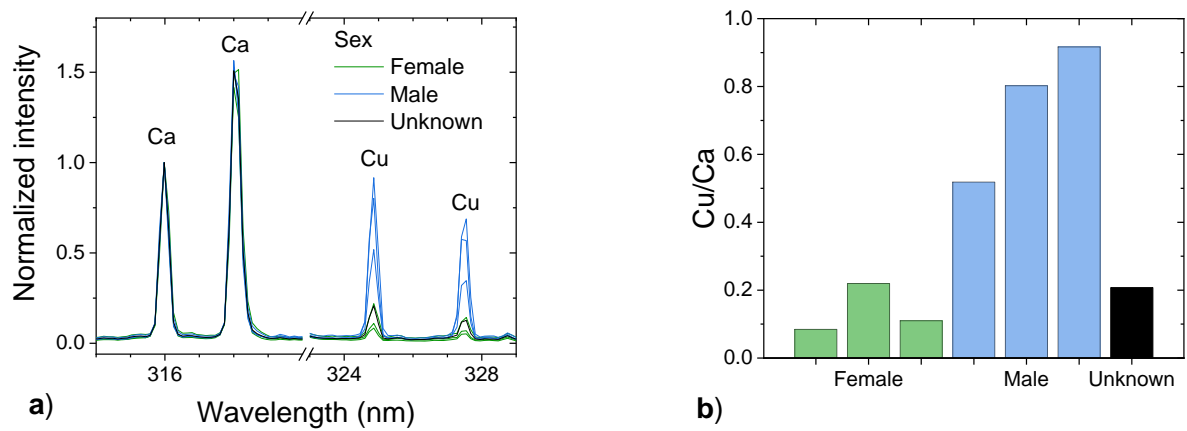

Fig. 5: Variation of Cu concentration in bones corresponding to identified sexes. a) Emission peaks of Cu normalised with respect to the Ca peak at 315.887 nm, in spectra recorded from the female bones (green), male bones (blue), and the bone of individual SK4.37 whose biological sex is unknown (black). b) Mean value of the Cu (324.754 nm)/Ca (315.887 nm) peak intensity ratio for each bone. Note that individual SK4.37 is within the range of the individuals identified as female.

**Tables**

Table 1: List of individuals and bone samples examined in this study.

| Individual | Skeleton ID | Sex | Bone element | Preservation[a] |
|---|---|---|---|---|
| 1 | SK 3.25 | Female | Femur | 2 |
| 2 | SK 3.50 | Male | Scapula Femur | 5+ |
| 3 | SK 4.36 | Male | Clavicle Ilium | 2 |
| 4 | SK 4.37 | Unknown | Tibia | 3 |
| 5 | SK 4.45 | Female | Humerus | 4 |

[a] Preservation reported in grades 0 (excellent) to 5+ (extremely poor) after McKinley *et al* (McKinley, 2004).

Table 2: Atomic emission lines in the LIBS spectra shown in Figure 2.

| Group | Element[a] | Wavelength (nm)[b] | |
|---|---|---|---|
| **1** | P I | 213.618, 213.618<br>214.914 | Biogenic |
| | Ca I | 452.694<br>457.855<br>458.140<br>458.587<br>458.596 | Biogenic |
| | Ca II | 315.887<br>318.052<br>318.128<br>409.710<br>410.982, 411.028 | |
| | Mg I<br>Mg II | 285.213<br>279.553<br>280.270 | |
| | Na I | 330.237, 330.298 | Biogenic |
| | Sr I | 407.771<br>460.733 | Biogenic |
| | Ba II | 455.403 | Biogenic |
| **2** | Al I | 308.215<br>309.271, 309.284 | Diagenic |
| | Fe I | 290-299 range<br>301.5-308 range<br>404.581<br>406.359<br>407.174 | Diagenic |
| | Mn I | 403.076<br>403.307, 403.449 | Diagenic |

[a] I: neutral atom, II : singly ionised atom.

[b] The emission lines and spectral ranges were identified based on the NIST atomic spectra database (Kramida, 2019) and have been used as input to the NN model.

Table 3: Data normalization methods tested in the present work.

| Normalization method | Equation | Success rate (%)[a] | Process time (s)[b] |
|---|---|---|---|
| Maximum intensity | $\dfrac{x_i}{max\,(x_i)}$ | 93.2 | 27.2 |
| Sum of the intensities | $\dfrac{x_i}{\Sigma(x_i)}$ | 96.4 | 68.8 |
| Square root of the sum of the squared intensities (Unit vector) | $\dfrac{x_i}{\sqrt{\Sigma(x_i^2)}}$ | 96.2 | 32.4 |
| Variance | $\sigma^2(x_i)$ | 94.4 | 166.2 |

[a] The SC values derived from the NN analysis of the test dataset

[b] The required process time that the NN method required to convergence.

Table 4: Spectrum correlation (SC) values derived from NN analysis of LIBS spectral data.[a]

| Group | Individuals | 1 | 2 | 3 | 4 | 5 | assigned to person |
|---|---|---|---|---|---|---|---|
| | | | | SC % | | | |
| 1 | 1 (2 bones) | 100 | 0 | 0 | 0 | 0 | 1 |
| | 2 (2 bones) | 0 | 100 | 0 | 0 | 0 | 2 |
| | 3 (1 bone) | 0 | 0 | 100 | 0 | 0 | 3 |
| | 4 (1 bone) | 0 | 0 | 0 | 100 | 0 | 4 |
| | 5 (1 bone) | 0 | 0 | 0 | 0 | 100 | 5 |
| 2 | 1 (2 bones) | 92 | 0 | 0 | 8 | 0 | 1 |
| | 2 (2 bones) | 6 | 94 | 0 | 0 | 0 | 2 |
| | 3 (1 bone) | 0 | 0 | 100 | 0 | 0 | 3 |
| | 4 (1 bone) | 0 | 18 | 73 | 0 | 9 | unassigned |
| | 5 (1 bone) | 0 | 0 | 0 | 0 | 100 | 5 |
| 1 & 2 | 1 (2 bones) | 100 | 0 | 0 | 0 | 0 | 1 |
| | 2 (2 bones) | 0 | 100 | 0 | 0 | 0 | 2 |
| | 3 (1 bone) | 14 | 0 | 86 | 0 | 0 | 3 |
| | 4 (1 bone) | 0 | 0 | 0 | 100 | 0 | 4 |
| | 5 (1 bone) | 0 | 0 | 0 | 0 | 100 | 5 |

[a] The NN analysis was performed on random samples.

Table 5: NN-LIBS analysis of spectral data for assigning two "unknown" bones. [a]

| Group | Individuals | 1 | 2 | 3 | 4 | 5 | assigned to person |
|---|---|---|---|---|---|---|---|
| | | | | SC % | | | |
| 1 | 1 (SK3.25 sca) | 0 | 0 | 34 | 53 | 13 | unassigned |
| | 2 (SK3.50 cla) | 0 | 91 | 0 | 0 | 9 | 2 |
| 2 | 1 (SK3.25 sca) | 0 | 0 | 72 | 10 | 18 | unassigned |
| | 2 (SK3.50 cla) | 14 | 69 | 2 | 1 | 14 | unassigned |
| 1 & 2 | 1 (SK3.25 sca) | 0 | 0 | 90 | 4 | 6 | 3 |
| | 2 (SK3.50 cla) | 3 | 82 | 2 | 0 | 15 | 2 |

[a] The NN analysis was performed to discriminate two bones that were characterized as unknown. The first test bone was from individual 1 and the second one from individual 2.